\documentstyle[12pt]{article}
\begin{document}
\begin{flushright}
hep-th/0111253\\
November 2001
\end{flushright}
\vskip 3cm
\begin{center}
{\bf \Large { Superfield approach to a novel symmetry\\
for non-Abelian gauge theory}}

\vskip 3cm

{\bf R.P.Malik}
\footnote{ E-mail address: malik@boson.bose.res.in  }\\
{\it S. N. Bose National Centre for Basic Sciences,} \\
{\it Block-JD, Sector-III, Salt Lake, Calcutta-700 098, India} \\

\vskip 3.5cm

\end{center}

\noindent
{\bf Abstract}:
In the framework of superfield formalism,
we demonstrate the existence of a new local, covariant,
continuous and nilpotent (dual-BRST) 
symmetry for the BRST invariant Lagrangian density of a self-interacting
two ($1 + 1$)-dimensional (2D) non-Abelian gauge theory 
(having no interaction with matter fields). The local and nilpotent
Noether conserved charges corresponding to the above 
continuous symmetries find their geometrical interpretation
as the translation generators along the odd (Grassmannian)  
directions of the four ($2 + 2)$-dimensional compact supermanifold. 
\baselineskip=16pt

%\vskip 1cm

\newpage

\noindent
{\bf 1 Introduction}

\noindent
For the covariant canonical quantization of the (non-)Abelian gauge theories
(endowed with the first-class constraints in the language of Dirac [1,2]), the
Becchi-Rouet-Stora-Tyutin (BRST) symmetry [3,4] plays a decisive role where 
unitarity and gauge invariance are respected together at any arbitrary order
of perturbation theory. The BRST formalism is indispensable in the context
of modern developments in topological field theories [5-7] and (super)string
theories (see, e.g., Ref.[8] and references therein). 
Its elegant extension to include second-class constraints in its folds [9], 
its mathematically consistent inclusion in the Batalin-Vilkovisky formalism 
[10,11], its geometrical interpretation in the framework of superfield 
formulation [12-16], etc., have elevated the subject of BRST formalism to 
a level where physical ideas (connected with the gauge theories) and 
underlying mathematical concepts (related to the differential geometry 
and cohomology) have merged together in such a fashion
that it has become an exciting and
interesting area of research for the past many years. 
Recently, in a set of papers [17-21], a possible connection between
the local, continuous and covariant symmetries and their generators on
the one hand and the de Rham cohomology operators
\footnote{ On an ordinary spacetime manifold without a boundary, the set
$(d, \delta, \Delta)$ (with $d = dx^\mu \partial_{\mu}, \delta = \pm * d *,
\Delta= d \delta + \delta d, * = $ Hodge duality operation) is known as
the de Rham cohomology operators of differential geometry where $(\delta) d$
and $\Delta$ are the (co)-exterior derivatives and the Laplacian operator
respectively. They obey an algebra: $ d^2 = \delta^2 = 0, [\Delta, d ] =
[\Delta, \delta ] = 0, \Delta = (\delta + d)^2$ showing that $\Delta$
is the Casimir operator.} of differential geometry
on the other hand, has been established in the Lagrangian formulation for
the case of 2D free- as well as interacting (non-)Abelian gauge theories.
Exploiting these symmetries, the topological nature of 2D free Abelian-
and self-interacting non-Abelian gauge theories (having no interaction with
matter fields) has been demonstrated [22]. The existence of such kind of local
symmetries has also been shown for the physical $(3+1)$-dimensional (4D)
free two-form Abelian gauge theory [23].

In the superfield approach [12-16] to BRST formalism for the 
$p$-form ($ p = 1, 2, 3.....$) gauge theories, the curvature ($(p+1)$-form)
tensor is restricted to be flat along the Grassmannian directions of the
$(D+2)$-dimensional supermanifold, parametrized by $D$- number of spacetime
(even) coordinates and two Grassmannian (odd) coordinates.
This flatness condition, popularly known as horizontality condition
\footnote{ This restriction has been referred to as the ``soul-flatness''
condition in Ref. [24] which amounts to setting the Grassmannian components
of the super curvature ($(p+1)$-form) tensor equal to zero.}, provides the 
origin for the existence of (anti-)BRST symmetry transformations and leads to 
the geometrical interpretation of the conserved and nilpotent ($Q_{b}^2 = 0, 
\bar Q_{b}^2 = 0$) (anti-)BRST charges ($\bar Q_{b})Q_{b}$ as the translation
generators along the Grassmannian directions of the $(D+2)$-dimensional
supermanifold. In this derivation, the super exterior derivative $\tilde d$
and the Maurer-Cartan equation for the definition of the curvature
tensor are exploited together for the imposition of the horzontality condition.
In a recent paper [25], all the three super de Rham cohomology operators,
defined on the $(2+2)$-dimensional supermanifold, have been exploited to
show the existence of (anti-)BRST- and (anti-)co-BRST 
symmetries as well as a bosonic symmetry for the case of a free 2D Abelian gauge
theory in the framework of superfield formulation. The logical explanation
for the existence of BRST- and co-BRST symmetries has been pointed out by 
exploiting the interplay between discrete and continuous symmetries [25].

The purpose of the present paper is to show the existence of a new local,
covariant, continuous and nilpotent symmetry for the 2D self-interacting 
non-Abelian gauge theory by exploiting the
mathematical power of the super co-exterior derivative ($ \tilde \delta$) of
the super de Rham cohomology operators. Here $\tilde \delta = \pm \star 
\tilde d \star$ is the super co-exterior derivative on the 
($2 + 2$)-dimensional supermanifold with $\star$ as the Hodge duality 
operation. We show that the imposition of an analogue of the horizontality
condition w.r.t. super cohomological operator $\tilde \delta$ 
leads to the derivation of (anti-)co-BRST symmetry which is exactly identical 
to the corresponding symmetry discussed
in the Lagrangian formulation {\it alone} [19,22]. As the (anti-)BRST charges 
$(\bar Q_{b})Q_{b}$ turn out to be the generators of translation along
the Grassmannian directions of the supermanifold, in a similar fashion, the
conserved and nilpotent ($Q_{d}^2 = \bar Q_{d}^2 = 0$) (anti-)co-BRST charges 
$(\bar Q_{d})Q_{d}$ find their geometrical interpretation as the translation
generators along the Grasmmannian directions of the ($2+2)$-dimensional
supermanifold. We also demonstrate that there is
a mapping between super cohomological operators and conserved charges
of the theory as: $ \tilde D = \tilde d + \tilde A \Leftrightarrow (Q_{b}, 
\bar Q_{b}),\tilde \delta \Leftrightarrow (Q_{d}, \bar Q_{d})$
where $\tilde D$ is the super covariant derivative and $\tilde A$ is the 
connection super one-form. Even though the anti-commutator of the (anti-)BRST-
and (anti-)co-BRST symmetry transformations is a bosonic symmetry for the
Lagrangian density, it is not possible to obtain this symmetry from the
anti-commutator of operators $\tilde D = \tilde d + \tilde A$ 
and $\tilde \delta$.
We argue about this problem and try to provide a mathematical reasoning for the
absence of this symmetry for the non-Abelian gauge theory 
in the framework of superfield formulation.

The material of our paper is organized as follows. In Sec. 2, we 
briefly recapitulate the key points of our earlier works on self-interacting
2D non-Abelian gauge theory [19,22] and set up the notations as well as 
conventions. Section 3 is devoted to a concise description of (anti-)BRST
symmetries in the framework of superfield formulation [14,15]. In Sec. 4,
we exploit the super operator $\tilde \delta$ (together with an analogue
of the horizontality condition w.r.t. this operator) for the derivation
of (anti-)co-BRST symmetries.  Finally, in Sec. 5, we make some concluding 
remarks and
discuss critically the mathematical reasons for the absence of a bosonic
symmetry for the non-Abelian gauge theory (that exists naturally in the Abelian
case [25] due to the Laplacian operator $\Delta = d \delta + \delta d$).\\

\noindent
{\bf 2 Preliminary: (anti-)BRST- and (anti-)co-BRST symmetries}

\noindent
Let us start off with the BRST invariant Lagrangian density ${\cal L}_{b}$
for the self-interacting two ($1 + 1)$-dimensional
\footnote{We follow here the conventions and notations such that the 2D flat
Minkowski metric is: $\eta_{\mu\nu} =$ diag $(+1, -1)$ and $\Box = 
\eta^{\mu\nu} \partial_{\mu} \partial_{\nu} = (\partial_{0})^2 - 
(\partial_{1})^2, \varepsilon_{\mu\nu} = - \varepsilon^{\mu\nu}, F^{a}_{01} 
= E^{a} = \partial_{0} A^{a}_{1} - \partial_{1} A^{a}_{0}
+ f^{abc} A^{b}_{0} A^{c}_{1}= F^{10a}, \varepsilon_{01} =
\varepsilon^{10} = + 1, \;D_{\mu} C^{a} = \partial_{\mu} C^{a} + f^{abc} 
A_{\mu}^{b} C^{c}, \alpha \cdot \beta = \alpha^{a} \beta^{a},
(\alpha \times \beta)^a = f^{abc} \alpha^{b} \beta^{c}$ where $\alpha$
and $\beta$ are the non-null vectors in the group space. Here the Greek 
indices: $\mu, \nu, \rho...= 0, 1$ correspond to spacetime directions.} 
non-Abelian gauge theory in the Feynman gauge [24,26-28]
$$
\begin{array}{lcl}
{\cal L}_{b} &=& - \frac{1}{4}\; F^{\mu\nu}\cdot F_{\mu\nu} 
- \frac{1}{2}\; (\partial_{\mu} A^{\mu}) \cdot (\partial_{\nu} A^{\nu})
- i \;\partial_{\mu} \bar C \cdot D^\mu C, \nonumber\\
&\equiv& \frac{1}{2}\; E \cdot  E 
- \frac{1}{2} \; (\partial_{\mu}  A^{\mu}) \cdot (\partial_{\nu} A^{\nu})
- i \;\partial_{\mu} \bar C \cdot D^\mu C, 
\end{array} \eqno(2.1)
$$
where $F^{a}_{\mu\nu} = \partial_{\mu} A_{\nu}^{a} -
\partial_{\nu} A_{\mu}^{a} + (A_{\mu} \times A_{\nu})^{a}$ is the field 
strength tensor derived from the one-form connection $ A = d x^\mu
A^{a}_{\mu} T^{a}$ by Maurer-Cartan equation $F = d A + A \wedge A$ 
with ($ F = \frac{1}{2} d x^\mu \wedge d x^\nu F_{\mu\nu}^{a} T^{a}$). Here
$T^{a}$ are the generators of the compact Lie algebra $ [ T^{a}, T^{b} ]
= f^{abc} T^{c}$ where $f^{abc}$ are the structure constants that can be
chosen to be totally antisymmetric in $ a, b, c$ 
(see, e.g., Ref.[28] for details). The Latin indices
$ a, b, c...= 1, 2, 3...$ correspond to the group indices (in the
colour space for the non-Abelian gauge theory).  The anti-commuting
($ (C^a)^ 2 = (\bar C^a)^2 = 0, C^a \bar C^b + \bar C^b C^a = 0$) (anti-)ghost
fields $(\bar C^a) C^a$ (which interact only with gauge fields $A_{\mu}^a$
in the loop diagrams) are required to be present in the theory to maintain the
unitarity and gauge invariance together. The auxiliary fields
$B^{a}$ and ${\cal B}^{a}$ can be introduced to linearize the gauge-fixing term
$-\frac{1}{2} [(\partial_{\mu} A^\mu)^{a}]^2$ (in the Feynman gauge) and
the kinetic energy term $\frac{1}{2}(E^{a})^2$ (because there is no magnetic 
component of $F_{\mu\nu}^{a}$ for the $(1+1)$-dimensional (2D) non-Abelian 
gauge theory) as 
$$
\begin{array}{lcl}
{\cal L}_{B} =  {\cal B}\cdot  E - \frac{1}{2} {\cal B} \cdot {\cal B}
+ B \cdot (\partial_{\mu}  A^{\mu}) + \frac{1}{2} B \cdot B
- i \partial_{\mu} \bar C \cdot D^\mu C. 
\end{array} \eqno(2.2)
$$
The above Lagrangian density
(2.1) respects the following off-shell nilpotent
$(s_{b}^2 = 0,  s_{d}^2 = 0)$ BRST ($s_{b}$)
\footnote{We adopt here the notations and conventions of Ref. [28]. In fact,
in its full glory, a nilpotent ($\delta_{B}^2 = 0$)
BRST transformation $\delta_{B}$ is equivalent to the product of an 
anti-commuting ($\eta C^a = - C^a \eta, \eta \bar C^a = - \bar C^a \eta$)
spacetime independent parameter $\eta$ and $s_{b}$ 
(i.e. $\delta_{B} = \eta \; s_{b}$) where $s_{b}^2 = 0$.} 
-and dual(co)-BRST ($s_{d}$) symmetry transformations [19,22]
$$
\begin{array}{lcl}
s_{b} A_{\mu} &=& D_{\mu} C, \qquad s_{b} C = -\frac{1}{2} C \times C, \qquad 
s_{b} \bar C = i B,  \nonumber\\
s_{b} {\cal B} &=& {\cal B} \times C, \qquad \;\; s_{b} B = 0, \;\;\qquad
\;s_{b} E = E \times C,
\end{array}\eqno(2.3)
$$
$$
\begin{array}{lcl}
s_{d} A_{\mu} &=& - \varepsilon_{\mu\nu} \partial^\nu \bar C, \qquad
s_{d} \bar C = 0, \qquad s_{d} C = - i {\cal B}, \nonumber\\
s_{d} {\cal B} &=& 0, \;\;\;\;\qquad s_{d} B = 0, 
\;\;\;\;\qquad s_{d} E = D_{\mu} \partial^{\mu} \bar C.
\end{array}\eqno(2.4)
$$
The anti-commutator of these nilpotent, local, continuous and covariant
symmetries (i.e. $ s_{w} = \{s_{b}, s_{d}\}$) leads to a bosonic symmetry
\footnote{This symmetry has {\it not} been discussed in Ref. [29] where the
nilpotent transformations (2.3) and (2.4) have been analyzed (without
indicating their possible connections with the de Rham cohomology operators)
on a compact Riemann surface. We thank Prof. N. Nakanishi for bringing
to our notice Ref. [29].}
$s_{w}$ ($ s_{w}^2 \neq 0$) transformations [19,22]
$$ 
\begin{array}{lcl}
s_{w} A_{\mu} &=&  D_{\mu} {\cal B}
+ \varepsilon_{\mu\nu} \partial^\nu B - i \varepsilon_{\mu\nu}
\partial^\nu \bar C \times C, 
 \quad s_{w} C = 0, \quad s_{w} \bar C = 0,  \nonumber\\
s_{w} (\partial_{\mu} A^\mu) &=& \partial_{\mu} D^\mu {\cal B}\;
+ \;i\; \varepsilon^{\mu\nu} \;\partial_{\mu} \bar C \times \partial_{\nu} C,
\;\;\;\;\qquad  \;\;\;\;\;\;\;s_{w} B\; =\; 0, \nonumber\\
s_{w} E &=& D_{\mu} (\partial^\mu \bar C \times C) - \varepsilon^{\mu\nu}
D_{\mu} D_{\nu} {\cal B} - D_{\mu} \partial^\mu B,
\quad s_{w} {\cal B} = 0,
\end{array} \eqno(2.5)
$$
under which the Lagrangian density (2.2) transforms to a total derivative 
\footnote{ The Lagrangian density (2.1) transforms to a total derivative
under the transformations: $ \tilde s_{w} A_{\mu}
= D_{\mu} E - \varepsilon_{\mu\nu} \partial^\nu (\partial_\rho A^\rho)
- i \varepsilon_{\mu\nu} \partial^\nu \bar C \times C, \tilde s_{w} C = 0,
\tilde s_{w} \bar C = 0$. This observation will be discussed in Sec.5.}.

Besides BRST- and co-BRST symmetry transformations (2.3) and (2.4), there
are anti-BRST- and anti-co-BRST symmetries that are also
present in the theory. To realize
these, one has to introduce another auxiliary field $\bar B$ 
(satisfying $ B + \bar B = i\; C \times \bar C$) to recast the
Lagrangian density (2.2) into the following forms [30]
$$
\begin{array}{lcl}
{\cal L}_{\bar B} =  {\cal B}\cdot  E - \frac{1}{2} {\cal B} \cdot {\cal B}
+ B \cdot (\partial_{\mu}  A^{\mu}) + \frac{1}{2} (B \cdot B + \bar B \cdot
\bar B) - i \partial_{\mu} \bar C \cdot D^\mu C, 
\end{array} \eqno(2.6a)
$$
$$
\begin{array}{lcl}
{\cal L}_{\bar B} =  {\cal B}\cdot  E - \frac{1}{2} {\cal B} \cdot {\cal B}
- \bar B \cdot (\partial_{\mu}  A^{\mu}) + \frac{1}{2} (B \cdot B + \bar B 
\cdot \bar B) - i D_{\mu} \bar C \cdot \partial^\mu C.
\end{array} \eqno(2.6b)
$$
The Lagrangian density (2.6b) transforms to a total derivative under the 
following off-shell nilpotent 
anti-BRST ($\bar s_{b}$)- and (anti-)co-BRST ($\bar s_{d}$) symmetry
transformations [19,22]
$$
\begin{array}{lcl}
\bar s_{b} A_{\mu} &=& D_{\mu} \bar C, \qquad \;\;\bar s_{b} \bar C 
= - \frac{1}{2} \bar C \times \bar C, \qquad \;\;
\bar s_{b}  C = i \bar B, \qquad \;\;\bar s_{b} \bar B = 0, \nonumber\\
\bar s_{b} E &=& E \times \bar C, \quad \bar s_{b} {\cal B} =
{\cal B} \times \bar C, \quad \bar s_{b} B = B \times \bar C,
\quad \bar s_{b} (\partial_{\mu} A^\mu) = \partial_{\mu} D^\mu \bar C,
\end{array} \eqno(2.8)
$$
$$
\begin{array}{lcl}
\bar s_{d} A_{\mu} &=& - \varepsilon_{\mu\nu} \partial^\nu C, \qquad
\bar s_{d}  C = 0, \qquad \bar s_{d} \bar C = + i {\cal B}, \qquad
\bar s_{d}  {\cal B} = 0, \nonumber\\
\bar s_{d} E &=& D_{\mu} \partial^\mu C, \qquad \bar s_{d} \bar B = 0,
\qquad \bar s_{d} B = 0, \qquad \bar s_{d} (\partial_\mu A^\mu) = 0.
\end{array} \eqno(2.9)
$$
The anti-commutator of these nilpotent symmetries leads to the transformations
equivalent to (2.5) (see, e.g., Ref. [19]). All the above continuous symmetry
transformations can be concisely expressed, in terms of the Noether
conserved charges $Q_{r}$ [19,22], as
$$
\begin{array}{lcl}
s_{r} \Psi = - i \; [ \Psi, Q_{r} ]_{\pm}, \qquad \;\;\;
Q_{r} = Q_{b}, \bar Q_{b}, Q_{d}, \bar Q_{d}, Q_{w}, Q_{g}, 
\end{array} \eqno(2.10)
$$
where brackets $[\;, \;]_{\pm}$ stand for the (anti-)commutators for 
any arbitrary generic field $\Psi$ being (fermionic)bosonic
in nature. Here the conserved ghost charge $Q_{g}$ generates
the continuous scale transformations: $ C \rightarrow e^{-\Sigma} C,
\bar C \rightarrow e^{\Sigma} \bar C, A_{\mu} \rightarrow A_{\mu},
B \rightarrow B, {\cal B} \rightarrow {\cal B}, \bar B \rightarrow \bar B$ 
(where $\Sigma$ is a global parameter). The local expression for $Q_{r}$
are given in Refs. [19,22].\\

\noindent
{\bf 3 Horizontality condition and (anti-)BRST symmetries}

\noindent
We begin here with a four ($2 + 2$)-dimensional compact supermanifold
parametrized by the superspace coordinates $Z^M = (x^\mu, \theta, \bar \theta)$
where $x^\mu (\mu = 0, 1)$ are the two even (bosonic) spacetime coordinates
and $\theta$ and $\bar \theta$ are the two odd (Grassmannian) coordinates
(with $\theta^2 = \bar \theta^2 = 0, 
\theta \bar \theta + \bar \theta \theta = 0)$. On this supermanifold, one can
define a supervector superfield $V_{s}$ with the following component
multiplet superfields [14-16]
$$
\begin{array}{lcl}
V_{s} = \Bigl ( B_{\mu} (x, \theta, \bar \theta), 
\;\Phi (x, \theta, \bar \theta),
\;\bar \Phi (x, \theta, \bar \theta) \Bigr ),
\end{array} \eqno(3.1)
$$
where the group valued superfields $B_{\mu} = B_{\mu}^a T^a, 
\Phi = \Phi^a T^a, \bar \Phi = \bar \Phi^a T^a$ can be expanded in terms
of the basic fields 
($A_\mu =A_\mu^aT^a, C=C^aT^a, \bar C=\bar C^a T^a$) and  auxiliary fields
($B=B^aT^a, \bar B=\bar B^aT^a, {\cal B}={\cal B}^aT^a$) of  
(2.6) and some extra secondary fields as follows
$$
\begin{array}{lcl}
(B_{\mu}^aT^a) (x, \theta, \bar \theta) &=& (A_{\mu}^aT^a) (x) 
+ \theta\; (\bar R_{\mu}^a T^a) (x) + \bar \theta\; (R_{\mu}^a T^a) (x) 
+ i \;\theta \;\bar \theta (S_{\mu}^aT^a) (x), \nonumber\\
(\Phi^aT^a)(x, \theta, \bar \theta) &=& (C^aT^a) (x) 
+ i\; \theta (\bar B^a T^a) (x)
+ i \;\bar \theta\; ({\cal B}^aT^a) (x) 
+ i\; \theta\; \bar \theta \;(s^a T^a) (x), \nonumber\\
(\bar \Phi^aT^a) (x, \theta, \bar \theta) &=& (\bar C^a T^a) (x) 
+ i \;\theta\;(\bar {\cal B}^aT^a) (x) + i\; \bar \theta \;(B^a T^a) (x) 
+ i \;\theta \;\bar \theta \;(\bar s^aT^a) (x).
\end{array} \eqno(3.2)
$$
The expansions are along the odd (fermionic) superspace coordinates 
$\theta$ and $\bar \theta$ and even (bosonic)  $(\theta \bar\theta)$ 
directions of the supermanifold.  All the fields are local functions
of spacetime coordinates $x^\mu$ alone (i.e.,$A_{\mu} (x,0,0) = A_\mu (x),
C^a (x,0,0) = C^a (x)$ etc.). It is straightforward to see that the local 
fields $ R_{\mu}^a (x), \bar R_{\mu}^a (x),
C^a (x), \bar C^a (x), s^a (x), \bar s^a (x)$ are fermionic (anti-commuting) 
in nature and the bosonic (commuting) local fields are: $A_{\mu}^a (x), 
S_{\mu}^a (x), {\cal B}^a (x), \bar {\cal B}^a (x),
B^a (x), \bar B^a (x)$ in the above expansion so that bosonic-
 and fermionic degrees of freedom can match. This requirement is essential
for the validity and sanctity of any arbitrary supersymmetric theory in the 
superfield formulation. In fact, all the secondary fields will be expressed 
in terms of basic fields due to the restrictions emerging from the application 
of ``horizontality'' condition on the super curvature (2-form) tensor 
$\tilde F$, defined through Maurer-Cartan equation, as
$$
\begin{array}{lcl} 
\tilde F =  \frac{1}{2}\; (d Z^M \wedge d Z^N)\;
\tilde F_{MN} = \tilde d \tilde A + \tilde A \wedge \tilde A \equiv
\tilde D \tilde A,
\end{array} \eqno(3.3)
$$
where covariant superderivative operator $\tilde D = \tilde d + \tilde
A $ is the generalization of the ordinary covariant derivative operator
$ D = d + A $.  Here super exterior derivative $\tilde d$ and 
connection super one-form $\tilde A$ are defined, in terms of
super differentials $dZ^M = (d x^\mu, d \theta, d \bar \theta)$ and component
superfields ($B_{\mu} = B_\mu^aT^a, \Phi = \Phi^aT^a,
\bar \Phi = \bar\Phi^aT^a$) of (3.1), as
$$
\begin{array}{lcl}
\tilde d &=& \;d Z^M \;\partial_{M} = d x^\mu\; \partial_\mu\;
+ \;d \theta \;\partial_{\theta}\; + \;d \bar \theta \;\partial_{\bar \theta},
\nonumber\\
\tilde A &=& d Z^M\; \tilde A_{M} = d x^\mu \;B_{\mu} (x , \theta, \bar \theta)
+ d \theta\; \bar \Phi (x, \theta, \bar \theta) + d \bar \theta\;
\Phi ( x, \theta, \bar \theta),
\end{array}\eqno(3.4)
$$
where partial derivatives, with respect to superspace coordinates, are
$$
\begin{array}{lcl}
\partial_{M} = {\displaystyle \frac{\partial}{\partial Z^M}},\quad
\partial_{\mu} = {\displaystyle \frac{\partial}{\partial x^\mu}},\quad
\partial_{\theta} = {\displaystyle \frac{\partial}{\partial \theta}},\quad
\partial_{\bar \theta} = {\displaystyle \frac{\partial}{\partial \bar \theta}}.
\end{array}\eqno (3.5)
$$
Now we impose the ``horizontality'' condition on $\tilde F$. 
Mathematically, this amounts to the imposition of
the following restriction [12-16]
$$
\begin{array}{lcl} 
\tilde F = \tilde d \tilde A + \tilde A \wedge \tilde A
\equiv d  A + A \wedge A = F \equiv D A,
\end{array} \eqno(3.6)
$$
where $ A \wedge A = \frac{1}{2} d x^\mu \wedge d x^\nu [ A_{\mu}, A_{\nu} ]
\equiv \frac{1}{2} d x^\mu \wedge d x^\nu (A_{\mu} \times A_{\nu})$. In words,
this requirement implies the ``flatness'' of all the components of the
super curvature (2-form) tensor $\tilde F_{MN}$ that are directed along the 
 $\theta$ and/or $\bar \theta$ directions of the supermanifold. More explicitly,
this restriction requires setting of the coefficients of $ d x^\mu \wedge
d \theta, d x^\mu \wedge d \bar \theta, d \theta \wedge d \bar \theta,
d \theta \wedge d \theta, d \bar \theta \wedge d \bar \theta$ equal to zero.
To achieve this, we expand the l.h.s. of (3.6) where the individual terms are
$$
\begin{array}{lcl}
\tilde d \tilde A &=& (d x^\mu \wedge d x^\nu)\;
(\partial_{\mu} B_\nu) - (d \theta \wedge d \theta)\; (\partial_{\theta}
\bar \Phi) + (d x^\mu \wedge d \bar \theta)
(\partial_{\mu} \Phi - \partial_{\bar \theta} B_{\mu}) \nonumber\\
&-& (d \theta \wedge d \bar \theta) (\partial_{\theta} \Phi 
+ \partial_{\bar \theta} \bar \Phi) 
+ (d x^\mu \wedge d \theta) (\partial_{\mu} \bar \Phi - \partial_{\theta}
B_{\mu}) - (d \bar \theta \wedge d \bar \theta)
(\partial_{\bar \theta} \Phi), 
\end{array}\eqno(3.7)
$$
$$
\begin{array}{lcl}
\tilde A \wedge \tilde A &=& (d x^\mu \wedge d x^\nu)\;
(B_{\mu} B_\nu) - (d \theta \wedge d \theta)\; (\bar \Phi
\bar \Phi) + (d x^\mu \wedge d \bar \theta)
(B_{\mu} \Phi - \Phi B_{\mu}) \nonumber\\
&-& (d \theta \wedge d \bar \theta) (\bar \Phi \Phi 
+  \Phi \bar \Phi) 
+ (d x^\mu \wedge d \theta) (B_{\mu} \bar \Phi - \bar \Phi
B_{\mu}) - (d \bar \theta \wedge d \bar \theta)
(\Phi \Phi).
\end{array}\eqno(3.8)
$$
Ultimately, the application of soul-flatness (horizontality) condition,
in its gory details, leads to the following relationships [14,15]
$$
\begin{array}{lcl}
R_{\mu} \;(x) &=& D_{\mu}\; C(x), \qquad 
\bar R_{\mu}\; (x) = D_{\mu}\;
\bar C (x), \qquad \;s\; (x) = (\bar B \times C) (x),
\nonumber\\
S_{\mu}\; (x) &=& D_{\mu} B\; (x) - i (D_{\mu} C \times \bar C)\;(x)
\equiv - D_{\mu} \bar B\; (x) + i (D_{\mu} \bar C \times  C)\; (x),
\nonumber\\
{\cal B}\; (x) &=& \frac{i}{2}\; (C \times C) (x), \quad
\bar {\cal B}\; (x) = \frac{i}{2}\;  (\bar C \times \bar C) (x),
 \quad \bar s\;(x) = - (B \times \bar C) (x), 
\end{array} \eqno(3.9)
$$
which demonstrates that all the secondary fields can be expressed in terms of
the basic- and auxiliary fields of the Lagrangian density (2.6). Besides
the above relations (3.9), the horizontality condition also leads to the 
systematic derivation of the relationship $ B + \bar B = i\; (C \times \bar C)$
[30] which results in from setting the coefficient of $(d\theta \wedge 
d\bar \theta$) equal to zero.

The insertion of all the above values for $R_{\mu}, \bar R_{\mu}, S_\mu,
{\cal B}, \bar {\cal B}, s , \bar s$
in the expansion (3.2) leads to
the derivation of (anti-)BRST symmetries for the non-Abelian gauge theory.
In addition, this exercise provides  the physical interpretation for the
(anti-)BRST charges as the generators of translations along the Grassmannian
directions of the supermanifold. Both these observations can be succinctly 
expressed, in a combined way, as
$$
\begin{array}{lcl}
B_{\mu}\; (x, \theta, \bar \theta) &=& A_{\mu} (x) 
+ \;\theta\; (\bar s_{b} A_{\mu} (x)) 
+ \;\bar \theta\; (s_{b} A_{\mu} (x)) 
+ \;\theta \;\bar \theta \;(s_{b} \bar s_{b} A_{\mu} (x)), \nonumber\\
\Phi\; (x, \theta, \bar \theta) &=& C (x) \;+ \; \theta\; (\bar s_{b} C (x))
\;+ \;\bar \theta\; (s_{b} C (x)) 
\;+ \;\theta \;\bar \theta \;(s_{b}\; \bar s_{b} C (x)), 
 \nonumber\\
\bar \Phi\; (x, \theta, \bar \theta) &=& \bar C (x) 
\;+ \;\theta\;(\bar s_{b} \bar C (x)) \;+\bar \theta\; (s_{b} \bar C (x))
\;+\;\theta\;\bar \theta \;(s_{b} \;\bar s_{b} \bar C (x)),
\end{array} \eqno(3.10)
$$
It will be noticed that in this interpretation equations (2.3)
(with $s_{b} \bar B = \bar B \times C$), (2.8) and 
(2.10) play very important role.
In fact, it is the mathematical power of the cohomological super operator
$\tilde d$ (along with the Maurer-Cartan equation)
that provides the geometrical interpretation for $Q_{b}$ and
$\bar Q_{b}$ as translation generators (cf.(2.10),(3.10)). Thus, the mapping is:
$ \tilde D = \tilde d + \tilde A \Leftrightarrow (Q_{b}, \bar Q_{b})$ 
but the ordinary exterior derivative $D = d + A$ (along with the 
Maurer-Cartan equation)
is identified with $Q_{b}$ {\it alone} because the latter
increases the ghost number of a state by one [17--21] as $D$ increases the 
degree of a form by one on which it operates. At this juncture, one noteworthy 
point is the fact that, after the imposition of the horizontality condition,
the 2-form super curvature $\tilde F (x, \theta, \bar \theta)$ 
for the Abelian gauge theory becomes
an ordinary 2-form curvature (i.e., $\tilde F (x,\theta,\bar\theta) 
= F (x)$) [25]. The same does not hold good in the case of the 
non-Abelian gauge 
theory. In fact, even after the imposition of the horizontality condition, 
the 2-form super curvature tensor $\tilde F = (dZ^M \wedge dZ^N) 
\tilde F_{MN} (x,\theta,\bar\theta)$ is:
$$
\begin{array}{lcl}
&&(dZ^M \wedge dZ^N) \tilde F_{MN} (x,\theta,\bar\theta) = 
(dx^\mu \wedge dx^\nu) \bigl \{F_{\mu\nu} (x)
+\; \theta \;(F_{\mu\nu} (x) \times \bar C (x))\nonumber\\ 
&&+ \;\bar \theta \;(F_{\mu\nu} (x) \times C (x))
+\; \theta\bar\theta\; \bigl [ i (F_{\mu\nu} (x) \times B (x)) +
(F_{\mu\nu}(x) \times \bar C(x)) \times C (x) \bigr ] \bigr \}.
\end{array} \eqno(3.11)
$$
However, the kinetic
energy term of the Lagrangian density (2.1) remains intact:
(i.e., $ - \frac{1}{4} \;F^{\mu\nu}(x) \cdot F_{\mu\nu}(x) = - \frac{1}{4}\;
\tilde F^{MN}(x, \theta, \bar \theta) \cdot 
\tilde F_{MN}(x, \theta, \bar \theta)$). This condition is trivially
satisfied in the case of Abelian gauge theory where $\tilde 
F (x,\theta,\bar \theta) = F (x)$ [25].\\

\noindent
{\bf 4 Analogue of horizontality condition and (anti-)co-BRST symmetries}

\noindent
It is obvious from equation (3.6) that, on an ordinary 2D Minowskian flat
spacetime manifold, the two-form $F = D A \equiv d A + A \wedge A$ 
(constructed from $ d = dx^\mu \partial_\mu$ and $ A = d x^\mu A_{\mu}^a T^a$)
defines the curvature tensor $F_{\mu\nu}^a = \partial_{\mu} A_{\nu}^a
- \partial_{\nu} A_{\mu}^a + (A_{\mu} \times A_{\nu})^a$ for the non-Abelian
gauge theory.  The operation of the 2D dual exterior derivative $\delta
= - * d *$ on the connection one-form $A$ (i.e., $\delta A = 
(\partial_\mu A^{\mu a} T^a$) leads to the definition of the gauge-fixing
term of the Lagrangian density (2.1). It is interesting to note that the
action of the operator $ \Omega = - * D *$ (constructed from covariant 
derivative $D = d + A $ and Hodge duality $*$ operation) on the 1-form $A$
(with $ * A = * (dx^\mu A_\mu) = \varepsilon^{\mu\nu} d x_{\nu} A_\mu$)
$$
\begin{array}{lcl}
\Omega\; A \equiv
- *\; D\; * A = - *\; ( d * A + A \wedge * A) = 
- * d * A = \partial_{\mu} A^{\mu a} T^a,
\end{array} \eqno(4.1)
$$
also leads to the derivation of the gauge-fixing term. Of course, here
we have used the fact that: $ * (dx^\mu) = \varepsilon^{\mu\nu}
(d x_\nu), * (dx^\mu \wedge dx^\nu) = \varepsilon^{\mu\nu}, 
(A_{\mu} \times A^{\mu})^a = f^{abc} A_{\mu}^b A^{\mu c} = 0$. This argument
persists with the super operators as well. Thus, for our all practical 
computations in this section, we shall concentrate on (super)operators
$(\tilde \delta) \delta$ and their operation on the connection super one-form
$(\tilde A) A$ for our discussion of analogue of the horizontality condition.
Here, for the case of the $(2 + 2)$-dimensional supermanifold, the 
super co-exterior derivative is: $ \tilde \delta 
= - \star \tilde d \star$. The Hodge duality $\star$ operation
on the super differentials $(d Z^M)$ and their wedge products 
$(d Z^M \wedge d Z^N), (dZ^M \wedge dZ^N \wedge dZ^P)$
etc., defined on this supermanifold, is
$$
\begin{array}{lcl}
&&\star\; (dx^\mu) = \varepsilon^{\mu\nu}\; 
(d x_\nu \wedge d\theta \wedge d \bar\theta)), \qquad \;
\star\; (d\theta) = 
\frac{1}{2!} \varepsilon^{\mu\nu}
(d x_\mu \wedge dx_\nu \wedge d \bar \theta), \nonumber\\
&&\star\; (d \bar \theta) = 
\frac{1}{2!} \varepsilon^{\mu\nu}
(d x_\mu \wedge dx_\nu \wedge d  \theta), \qquad\;
\star\; (d x^\mu \wedge d x^\nu) = \varepsilon^{\mu\nu}
(d\theta \wedge d\bar\theta), \nonumber\\
&&\star\; (dx^\mu \wedge d \theta) = \varepsilon^{\mu\nu}
(dx_\nu \wedge d \bar\theta), \;\;\;\qquad\;\;\;
\star\; (dx^\mu \wedge d \bar \theta) = \varepsilon^{\mu\nu}
(dx_\nu \wedge d\theta), \nonumber\\
&&\star \; (d \theta \wedge d \theta) = 
\frac{1}{2!} \;
s^{\theta\theta}\;
\varepsilon^{\mu\nu}
(d x_\mu \wedge dx_\nu), \quad
\star \; (d \theta \wedge d \bar \theta) = 
\frac{1}{2!}\;
s^{\theta \bar \theta}\;\varepsilon^{\mu\nu}
(d x_\mu \wedge dx_\nu), \nonumber\\
&&\star \; (d \bar \theta \wedge d \bar \theta) = 
\frac{1}{2!}\;s^{\bar \theta \bar \theta}\;
 \varepsilon^{\mu\nu}\;(d x_\mu \wedge dx_\nu), \qquad\;
\star\; (dx_\mu \wedge d\theta \wedge d \bar\theta) =
\varepsilon_{\mu\nu}\; (dx^\nu), \nonumber\\
&& \star\; (dx_\mu \wedge dx_\nu \wedge d\theta \wedge d\bar\theta) =
\varepsilon_{\mu\nu}, \qquad
\star\; (dx_\mu \wedge dx_\nu \wedge d \theta) =
\varepsilon_{\mu\nu}\; (d\bar\theta),\nonumber\\
&&\star\; (dx_\mu \wedge dx_\nu \wedge d \bar\theta) =
\varepsilon_{\mu\nu}\; (d\theta),\qquad
\star\;(dx_\mu \wedge dx_\nu \wedge d \theta \wedge d\theta)
= \varepsilon_{\mu\nu}\; s^{\theta\theta},
\end{array} \eqno(4.2)
$$
where $s's$ are symmetric (e.g., $s^{\theta \bar \theta} 
= s^{\bar\theta\theta}$ etc.). In the above we have
collected only a few of the $\star$ operations. The other $\star$
operations can be computed in an analogous mannar. With these as the backdrop,
we obtain the expression for the superscalar superfield 
$\tilde \delta \tilde A = - \star \tilde d \star \tilde A$ as
$$
\begin{array}{lcl}
\tilde \delta \tilde A = (\partial_\mu B^\mu) + s^{\theta\theta} 
(\partial_{\theta} \Phi) + s^{\bar \theta \bar\theta} (\partial_{\bar\theta}
\bar \Phi) + s^{\theta \bar \theta} (\partial_{\theta} \bar \Phi +
\partial_{\bar \theta} \Phi).
\end{array} \eqno(4.3)
$$
It will be noted that we have dropped all the terms in the computation of
($\tilde d \star \tilde A$) which contain (i) more than two differentials
in spacetime, and (ii) more than two differentials in Grassmannian variables.
After this only, we have applied another $\star$ operation on it.
Now we exploit the analogue of the horizontality condition w.r.t. 
$\tilde \delta$ (i.e., $\tilde \delta \tilde A = \delta A $) which is
nothing but equating the r.h.s. of equations (4.1) and (4.3). In other words, 
we set the coefficients of $s^{\theta\theta}, s^{\bar\theta\bar\theta},
s^{\theta\bar\theta}$
equal to zero. The ensuing restrictions on the superfields are
$$
\begin{array}{lcl}
\partial_{\theta} \bar \Phi + \partial_{\bar \theta} \Phi = 0, \quad\;\;
\partial_{\theta} \Phi = 0, \quad\;\;  
\partial_{\bar \theta} \bar \Phi = 0.
\end{array} \eqno(4.4)
$$
Exploiting the expansions (3.2), it can be checked that the above restrictions 
lead to
$$
\begin{array}{lcl}
s^a (x) = 0,  \quad \bar s^a (x) = 0, 
\quad \;{\cal B}^a (x) + \bar {\cal B}^a (x) = 0, 
\quad B^a (x) = \bar B^a (x) = 0,
\end{array} \eqno(4.5)
$$
and the following conditions on the component fields of
$B_{\mu} (x,\theta,\bar\theta)$
$$
\begin{array}{lcl}
\partial \cdot \bar R = 0, \qquad \partial \cdot R = 0, \qquad
\partial \cdot S = 0.
\end{array} \eqno(4.6)
$$
In terms of solutions (4.5) and
$R_\mu^a = - \varepsilon_{\mu\nu} \partial^\nu \bar C^a, 
\bar R_\mu^a = - \varepsilon_{\mu\nu} \partial^\nu  C^a, 
S_\mu^a = - \varepsilon_{\mu\nu} \partial^\nu {\cal B}^a$, 
the superfield expansion (3.2) can be 
re-expressed as
$$
\begin{array}{lcl}
B_{\mu}\; (x, \theta, \bar \theta) &=& A_{\mu} (x) 
- \;\theta\; \varepsilon_{\mu\nu}\;\partial^{\nu}  C (x) 
- \;\bar \theta\;\varepsilon_{\mu\nu}\; \partial^{\nu} \bar C (x) 
- \;i \;\theta \;\bar \theta \;
\varepsilon_{\mu\nu}\;\partial^{\nu}\; {\cal B} (x), \nonumber\\
\Phi\; (x, \theta, \bar \theta) &=& C (x) 
+ \;i \;\bar \theta\; {\cal B} (x), \;\;\qquad\;\;
\bar \Phi\; (x, \theta, \bar \theta) = \bar C (x) 
-\; i \; \theta\; {\cal B} (x).
\end{array} \eqno(4.7)
$$
Exploiting equations  (2.4) and (2.9), we can write the
above expansion in a form similar to (3.10). The resulting expansion,
in terms of (anti-)co-BRST symmetries, is
$$
\begin{array}{lcl}
B_{\mu}\; (x, \theta, \bar \theta) &=& A_{\mu} (x) 
+ \;\theta\; (\bar s_{d} A_{\mu} (x)) + \;\bar \theta\; (s_{d} A_{\mu} (x)) 
+ \;\theta \;\bar \theta \;(\bar s_{d} s_{d} A_{\mu} (x)), \nonumber\\
\Phi\; (x, \theta, \bar \theta) &=& C (x) - \; \bar \theta\; (s_{d} C (x)),
\qquad \;\; \bar \Phi\; (x, \theta, \bar \theta) = \bar C (x) 
- \theta\; (\bar s_{d} \bar C (x)).
\end{array} \eqno(4.8)
$$
At this stage, it is interesting to compare and contrast the finer details
of the expansion (3.10) and (4.7) which are connected with (anti-)BRST-
and (anti-)co-BRST symmetries. We pin-point the facts that:
(i) the (anti-)BRST and (anti-)co-BRST symmetry
transformations are generated along the $\theta (\bar \theta)$ directions
of the supermanifold. (ii) The translation generators along the Grassmannian
directions of the supermanifold are the conserved and nilpotent
(anti-)BRST- and (anti-)co-BRST charges (cf.(2.10)). (iii) 
For the odd (fermionic) superfields, the
translations are either along $\theta$ or $\bar \theta$ directions 
for the case of (anti-)co-BRST symmetries. 
This is not the case with (anti-)BRST symmetries (cf.(3.10)).
(iv) For the  bosonic superfield, the translations are along both $\theta$ 
as well as $\bar \theta$ directions when we consider (anti-)BRST- and/or
(anti-)co-BRST symmetries. (v) Comparison between (3.10) and (4.7) shows
that the (anti-)BRST transformations are along $(\theta)\bar \theta$
directions for the odd fields $(C)\bar C$. {\it On the contrary, the 
(anti-)co-BRST transformations are the other way around}. (vi) The restrictions 
$ \tilde \delta \tilde A = \delta A$ and $\tilde D \tilde A = D A$
produce (anti-)co-BRST- and (anti-)BRST symmetry transformations. 
(vii) Form of the solutions (4.5) are such that it turns out to be the
{\it straightforward} generalization of the (anti-)co-BRST symmetries for
the Abelian gauge theory [25] as there are no structure constants $f^{abc}$
anywhere in (4.5). In contrast, the form of the solutions in (3.9) do
reflect the {\it nontrivial} generalization of the Abelian transformations 
[25] to the non-Abelian (anti-)BRST transformations. (viii) The expresions
for $R_\mu$ and $\bar R_\mu$ in (3.9) and (4.5) are such that the kinetic
energy- and gauge-fixing terms of (2.1) remain invariant under (anti-)BRST-
and (anti-)co-BRST symmetries, respectively.

For the (anti-)co-BRST symmetries
 the mapping is: $ \tilde \delta \Leftrightarrow (Q_{d},\bar Q_{d})$ but the 
ordinary co-exterior derivative $\delta$ is identified with
$Q_{d}$ {\it alone} because it decreases the ghost number 
of a state by one [19,22] as $\delta$ reduces the degree of a given form 
by one on which it operates.\\

\noindent
{\bf 5 Discussion}

\noindent
We have shown the existence of the local, covariant, continuous and nilpotent 
(anti-)BRST- and 
(anti-)co-BRST symmetries by exploiting the mathematical power of super
operators $\tilde D = \tilde d + \tilde A$ and $\tilde \delta$ 
together with the ldea of the generalized version of the ``horizontality''
condition [12-16]. In the framework of Lagrangian density alone,
these symmetry transformations have been obtained in equations (2.3), (2.4),
(2.8) and (2.9). In fact, there are six 
local, covariant, and continuous symmetry transformations in the theory
which have been discussed in Sec. 2. They obey the following algebra:
$$
\begin{array}{lcl}
&& s_{b}^2 = s_{d}^2 = \bar s_{b}^2 = \bar s_{d}^2 = 0, \qquad
s_{w} = \{ s_{b}, s_{d} \} = \{ \bar s_{b}, \bar s_{d} \}, \nonumber\\
&& s_{d} \bar s_{d} + \bar s_{d} s_{d} = 0, \quad
 s_{b} \bar s_{b} + \bar s_{b} s_{b} = 0, \quad
[s_{w}, s_{r} ] = 0, \quad s_{r} = s_{b}, \bar s_{b}, s_{d}, \bar s_{d},
s_{g}, \nonumber\\
&&i [s_{g}, s_{b}] = + s_{b}, \quad i [s_{g}, s_{d}] = - s_{d},\quad
i [s_{g}, \bar s_{b}] = - \bar s_{b}, \quad i [s_{g}, \bar s_{d}] 
= + \bar s_{d},
\end{array} \eqno(5.1)
$$
which is reminiscent of the algebra obeyed by the ordinary
de Rham cohomology operators [31-34] of differential geometry
(see, e.g., one of the foot-notes in Sec.1) defined on the 
flat ordinary compact 
manifold. In fact, for the free Abelian (two-dimensional  one-form
as well as four-dimensional two-form) gauge theories, it has been 
shown [17,18,22,23]
that there is one-to-one correspondence between local symmetry 
transformations for the Lagrangian density (and corresponding generators) on 
the one hand and the de Rham cohomology operators
$(d, \delta, \Delta)$
on the other hand. The existence of a discrete symmetry for the Lagrangian
density has been shown to correspond to the Hodge $(*)$ duality operation of
the differential geometry. In our recent work [25], we have exploited the
mathematical power of super de Rham cohomology operators $(\tilde d, 
\tilde \delta, \tilde \Delta=\tilde d \tilde \delta
+ \tilde \delta \tilde d)$ to demonstrate the geometrical origin
for the (anti-) BRST- and (anti-)co-BRST charges as the translation generators
along the Grassmannian directions of the $(2+2)$-dimensional supermanifold.
A bosonic symmetry (which is equivalent to the anti-commutator of these two
nilpotent symmetries) is shown to correspond to the translation along the
bosonic $\theta \bar \theta$-direction 
(equivalent to a couple of intertwined Grassmannian directions)
of the supermanifold. The bosonic generator for this symmetry
is shown to owe its origin to the Laplacian operator. In fact, the Hodge
decomposed versions for the 2D vector fields (i.e.,$R_{\mu}, \bar R_{\mu},
S_{\mu}$) emerge very naturally when we exploit the Laplacian operator to
show the existence of the bosonic symmetry by requiring the fermionic
fields ($R_{\mu}, \bar R_{\mu}, s, \bar s$)-
as well as auxiliary fields ($B, \bar B, {\cal B}, \bar {\cal B}$) 
to be zero
in the expansion (3.2).  This trick does not work in the case of non-Abelian 
gauge theory because the mathematical operators (we are dealing with) are:
$ D =  d +  A$ and $\Omega = - *  D * \equiv \delta$. These operators do
not form a closed algebra. In contrast, for the Abelian case, the de Rham
cohomology operators ($d, \delta, \Delta$) do close among themselves.
In addition, there is no analogue of the Hodge duality $(*)$ operation
as a discrete symmetry transformation for the case of
non-Abelian gauge theory. Thus, even though the 2D 
self-interacting non-Abelian gauge theory is a topological field theory [22]
(like 2D free Abelian gauge theory),
it is not an exact field theoretical model for the Hodge theory. In fact,
when we exploit the analogue of the horizontality condition w.r.t. the super
operator $\tilde \delta \tilde D + \tilde D \tilde \delta$, namely;
$$
\begin{array}{lcl}
\bigl ( \tilde \delta \tilde D + \tilde D \tilde \delta \bigr ) \tilde A
= \bigl (  \delta  D +  D  \delta \bigr ) A
\equiv  d x^\rho \; \bigl [D_\rho (\partial_\mu A^\mu) -
\varepsilon_{\rho\sigma} \partial^\sigma E \bigr ),
\end{array} \eqno(5.2)
$$
we do not obtain the Hodge decomposed version for the 2D vector fields
($R_\mu, \bar R_\mu, S_\mu$). Furthermore,
 neither of the bosonic symmetries $s_{w}$
(or its on-shell equivalent $\tilde s_{w}$) emerge from the above condition.
In fact, it can be also clearly seen that the following horizontality
restrictions with operators $\tilde \Omega = - \star \tilde D \star$
and $\tilde D = \tilde d + \tilde A$
$$
\begin{array}{lcl}
\bigl ( \tilde \Omega \tilde D + \tilde D \tilde \Omega \bigr ) \tilde A
= \bigl (  \Omega  D +  D  \Omega \bigr ) A
\equiv  d x^\rho \; \bigl [D_\rho (\partial_\mu A^\mu) -
\varepsilon_{\rho\sigma} D^\sigma E \bigr ),
\end{array} \eqno(5.3)
$$
do not lead to the derivation of the bosonic symmetry $s_{w}$ (or
its on-shell equivalent $\tilde s_{w}$) by invoking the requirements:
$ R_\mu = \bar R_\mu = s = \bar s = 0$ and $B = \bar B = {\cal B} =
\bar {\cal B} = 0$. On the contrary, for the 2D free Abelian one-form
gauge theory [25], we have demonstrated that the analogue of the horizontality
condition $ \tilde \Delta \tilde A = \Delta A$ leads to the 
derivation of the on-shell version $\tilde s_{w}$
(of the bosonic symmetry $s_{w}$) which is consistent with the Hodge
decomposition of the 2D vectors $R_\mu$ and $\bar R_\mu$. 

At the moment, it is an open problem to find out 
the mathematical resolution for the existence of the bosonic
symmetry $s_{w}$ (or its equivalent $\tilde s_{w}$) for the 2D non-Abelian
gauge theory. It will also be important to express the topological features
of these 2D theories in the language of the geometrical superfield 
formulation. These are some of the issues that are under investigation and
our results will be reported elsewhere [35].\\

\end{document}